\documentclass[a4paper,UKenglish]{lipics-v2016}
 
\usepackage{amsfonts}

\newcommand{\set}[1]{\left\{ #1\right\}}
\newcommand{\gilt}{:}
\newcommand{\sodass}{\,:\,}
\newcommand{\setGilt}[2]{\left\{ #1\sodass #2\right\}}

\newcommand{\realrange}[2]{\left[#1, #2\right]}

\newcommand{\unitrange}[2]{\realrange{0}{1}}

\newcommand{\llabel}[1]{\label{\labelprefix:#1}}
\newcommand{\labelprefix}{} 

\newcommand{\discussionsize}{\small}

\newcommand{\frage}[1]{}

\newenvironment{code}{\noindent
\begin{tabbing}%
\hspace{2em}\=\hspace{2em}\=\hspace{2em}\=\hspace{2em}\=\hspace{2em}\=%
\hspace{2em}\=\hspace{2em}\=\hspace{2em}\=\hspace{2em}\=\hspace{2em}\=%
\kill}{\end{tabbing}}

\newcommand{\labelcommand}{}
\newcommand{\captiontext}{}
\newsavebox{\codeparam}
\newcounter{lineNumber}
\newenvironment{disscodepos}[3]{%
\renewcommand{\labelcommand}{#2}%
\renewcommand{\captiontext}{#3}%
\sbox{\codeparam}{\parbox{\textwidth}{#3}}%
\begin{figure}[#1]\begin{center}\begin{code}\setcounter{lineNumber}{1}}{%
\end{code}\end{center}\caption{\llabel{\labelcommand}\captiontext}\end{figure}}

{\end{disscodepos}}

\newcommand{\Is}       {:=}

\newdimen\endofsize\endofsize=0.5em
\def\endofbeweis{~\quad\hglue\hsize minus\hsize
                 \hbox{\vrule height \endofsize width
\endofsize}\par}

\usepackage{microtype}
\usepackage{wrapfig}
\usepackage{tikz}

\def\MdR{\ensuremath{\mathbb{R}}}

\newcommand{\csch}[1]{{\color{blue}[CS: #1]}}
\renewcommand{\csch}[1]{}

\newcommand{\ie}{i.e.\ }
\newcommand{\etal}{et~al.~}
\newcommand{\eg}{e.g.\ }

\title{Optimal Longest Paths by Dynamic Programming\footnote{This work was partially supported by DFG grants SA 933/11-1.}}
\titlerunning{Optimal Longest Paths by Dynamic Programming} 

\author[1]{Tom\'a\v{s} Balyo}
\author[1]{Kai Fieger}
\author[1,2]{Christian Schulz}
\affil[1]{Karlsruhe Institute of Technology, Karlsruhe, Germany\\
  \texttt{\{tomas.balyo, christian.schulz\}@kit.edu}, fieger@ira.uka.de}
\affil[2]{University of Vienna, Vienna, Austria\\
  \texttt{christian.schulz@univie.ac.at}}

\authorrunning{T. Balyo et. al.} 

\Copyright{Tomas Balyo, Kai Fieger, Christian Schulz}


\EventLongTitle{16th International Symposium on Experimental Algorithms (SEA 2017)}
\EventShortTitle{SEA 2017}
\EventAcronym{SEA}
\EventYear{2017}
\EventDate{June 21--23, 2017}
\EventLocation{London, United Kingdom}
\EventLogo{}

\begin{document}

\maketitle

\begin{abstract}
We propose an optimal algorithm for solving the longest path problem in undirected weighted graphs. 
By using graph partitioning and dynamic programming, we obtain an algorithm 
that is significantly faster than other state-of-the-art methods. This enables us to solve instances that have previously been unsolved.
 \end{abstract}

\section{Introduction} 
The longest path problem (LP) is to find a simple path 
of maximum length between two given vertices of a graph 
where length is defined as the number of edges or the total weight of the edges in the path.
The problem is known to be NP-complete~\cite{NP} and has several applications such as designing circuit boards~\cite{Circuit1,Circuit2}, project planning~\cite{Brucker}, information retrieval~\cite{Wong} or patrolling algorithms for multiple robots in graphs~\cite{Multirobot}. 
For example, when designing circuit boards where the length difference between wires has to be 
kept small~\cite{Circuit1,Circuit2}, the longest path problem manifests itself when the length of shorter wires is supposed to be increased. 
Another example application is project planning/scheduling where the problem can be used to determine the least 
amount of time that a project could be completed in~\cite{Brucker}. 

We organize the rest of paper as follows.
After introducing basic concepts and  related work in Section~\ref{s:basicconcepts}, we present our \emph{main contribution} in Section~\ref{s:maincontribution}: an optimal algorithm for the longest path problem in undirected graphs. 
The main ingredient of our algorithm is a dynamic programming technique based on hierarchical partitions of the graph. 
A summary of extensive experiments done to tune the algorithm and evaluate its performance is presented in Section~\ref{s:experiments}. 
This includes a study of the algorithm's performance with respect to different partitioning strategies in order to find a good balance between the time spent
for partitioning and the overall runtime of our algorithm.
We also compare our algorithm with optimal algorithms presented in recent literature~\cite{stern}.
Experiments show that our new algorithm solves significantly more instances and is also faster than other optimal algorithms on \emph{large/hard} instances. 
Finally, we conclude with~Section~\ref{s:conclusion}.
\vfill
\pagebreak
\section{Preliminaries}
\label{s:basicconcepts}
\subsection{Definitions}
In the following we consider an undirected graph $G=(V,E,\omega)$ 
with edge weights $\omega: E \to \MdR_{\geq 0}$, $n = |V|$, and $m = |E|$.
We extend $\omega$ to sets, i.e.,
$\omega(E')\Is \sum_{e\in E'}\omega(e)$.
$N(v)\Is \setGilt{u}{\set{v,u}\in E}$ denotes the neighbors of $v$.
A \emph{subgraph} is a graph whose vertex and edge set are subsets of another graph. We call a subgraph induced if it has every possible edge.  
A subset of a graph's vertices is called a \emph{clique} if the graph contains an edge between every two distinct vertex of the subset.
A \emph{matching} is a subset of the edges of a graph where no two edges have vertices in common.
A sequence of vertices $s \to \cdots \to t$ such that each pair of consecutive vertices is connected by an edge, is called an \emph{$s$-$t$ path}. 
We say that $s$ is the source and $t$ is the target. 
A path is called \emph{simple} if it does not contain a vertex more than once.
The length of a path is defined by the sum of its edge weights. 
If the graph is unweighted, then edge weights are assumed to be one.

Given a graph $G=(V,E,\omega)$ as well as two vertices $s, t \in V$, the \emph{longest path} (LP) problem is to find the longest simple path from $s$ to $t$. 
Another version of the longest path problem is to find the longest simple path between any two vertices. 
However, the problem can be solved by introducing two vertices $s, t$, connecting them to all other vertices in the graph by edges of weight zero and then running algorithms tackling the LP problem~on~the~modified~instance.

A $k$-way partition of a graph is a division of $V$ into \emph{blocks} of vertices $V_1$,\ldots,$V_k$, \ie $V_1\cup\cdots\cup V_k=V$ and $V_i\cap V_j=\emptyset$
for $i\neq j$.
A \emph{balancing constraint} demands that 
$\forall i\in \{1..k\}\gilt |V_i|\leq L_{\max} := (1+\epsilon)\lceil\frac{|V|}{k}\rceil$
for some imbalance parameter $\epsilon$. 
The objective is typically to minimize the total \emph{cut} $\sum_{i<j}w(E_{ij})$ where 
$E_{ij}\Is\setGilt{\set{u,v}\in E}{u\in V_i,v\in V_j}$.

\subsection{Related Work}
This paper is based on the bachelor thesis of Kai Fieger \cite{thesisFieger}. 
Previous work by Stern \etal\cite{stern} mainly 
focuses on the possibility of applying algorithms that are usually used to 
solve the shortest path problem (SP) to the longest path problem. 
Stern \etal\cite{stern} make clear why LP is so difficult compared to SP. 
Several algorithms are presented that are frequently used to solve SP or other minimization search problem.
They are modified in order to be able to solve LP. 
The search algorithms are put into three categories: heuristic, uninformed and suboptimal. 
Each of the algorithms in the first two categories yields optimal solutions to the problem.
The most relevant category for this paper is heuristic searches. Here, a heuristic can provide 
extra information about the graph or the type of graph. Heuristic searches 
 require a heuristic function that estimates the remaining 
length of a solution from a given vertex of the graph. 
This can give important information helping to prune the search space. 
Stern \etal\cite{stern} show that heuristic searches can be used efficiently for the longest path problem. 
Some examples of algorithms in this category are Depth-First-Branch-and-Bound (DFBnB) and A*.
Another category represents ``uninformed'' searches, which do not require any information other 
than what is already given in the definition of the problem. 
Examples from this category are Dijksra's algorithm or DFBnB without a heuristic. Modifying 
these algorithms to fit LP basically leads to brute force algorithms, which means 
that they still have to look at every possible path in the search space. 
Hence, these uninformed search strategies are not very beneficial for LP. 
The last category are the suboptimal searches. The authors looked at a large 
number of these algorithms that only find approximations of a longest path. 
They are, however, not relevant for this paper since we present an optimal algorithm.
We are not aware of any recent work other than \cite{stern} related to optimal LP solving.
\paragraph*{Karlsruhe High Quality Partitioning}
Within this work, we use the open source multilevel graph partitioning framework KaHIP~\cite{kaffpa,kabapeE} (Karlsruhe High Quality Partitioning).
More precisely, we use the distributed evolutionary algorithm KaFFPaE contained therein to create partitions of our graphs that are better suited for our dynamic programming approach. 
The algorithms in KaHIP have been able to improve the best known partitioning results in the Walshaw Benchmark~\cite{soper2004combined} for many inputs using a short amount of time to create the partitions.
We refer the reader to \cite{kaffpaE} for more details.

\section{Longest Path by Dynamic Programming}
\label{s:maincontribution}
\label{sec:algo}

We now introduce the main contribution of our paper which is a new algorithm to tackle the longest path problem based on principles of dynamic programming. 
Hence, our algorithm is called ``Longest Path by Dynamic Programming'' (LPDP). 
Our algorithm solves the longest path problem (LP) for weighted undirected graphs. We start this section by introducing the main approach which includes preprocessing the graph as well as combining paths. At the end of the section, we show how to improve the approach by using hierarchical partitions of the input instance
and by reducing the number of auxiliary nodes which reduces the size of the search space.

\subsection{The Basic Approach}
\label{sec:approach}
A simple way to solve the longest path problem is \emph{exhaustive depth-first search}~\cite{stern}. 
A regular depth-first search (DFS) starts at a root vertex. By default, a vertex has two states: marked and unmarked. Initially, all vertices are unmarked, except the root which is marked.
DFS calls itself recursively on each unmarked vertex 
reachable by an edge from the current vertex, which is the parent of these vertices. The vertex is marked. 
Once it is done it backtracks to its parent. The search is finished once DFS backtracks at the root vertex.

Exhaustive DFS is a DFS that unmarks a vertex upon backtracking. In that way every simple path 
in the graph starting from the root vertex is explored. The LP problem can be solved with 
exhaustive DFS by using the start vertex as root. During the search the 
length of the current path is stored and compared to the previous best 
solution each time the target vertex is reached. If the current length is greater than 
that of the best solution, it is updated accordingly. When the search is done 
a path with maximum length from $s$ to $t$ has been found.
The \emph{main idea} of LPDP is to partition a graph into multiple blocks and run a search similar to 
exhaustive DFS on each block in a preprocessing step. Afterwards the results can be combined into a single longest path for $G$.

\subsubsection*{Partitioning and Preprocessing}
\label{sec:approach:step1}
\begin{figure}
  \centerline{\includegraphics[width=1\linewidth]{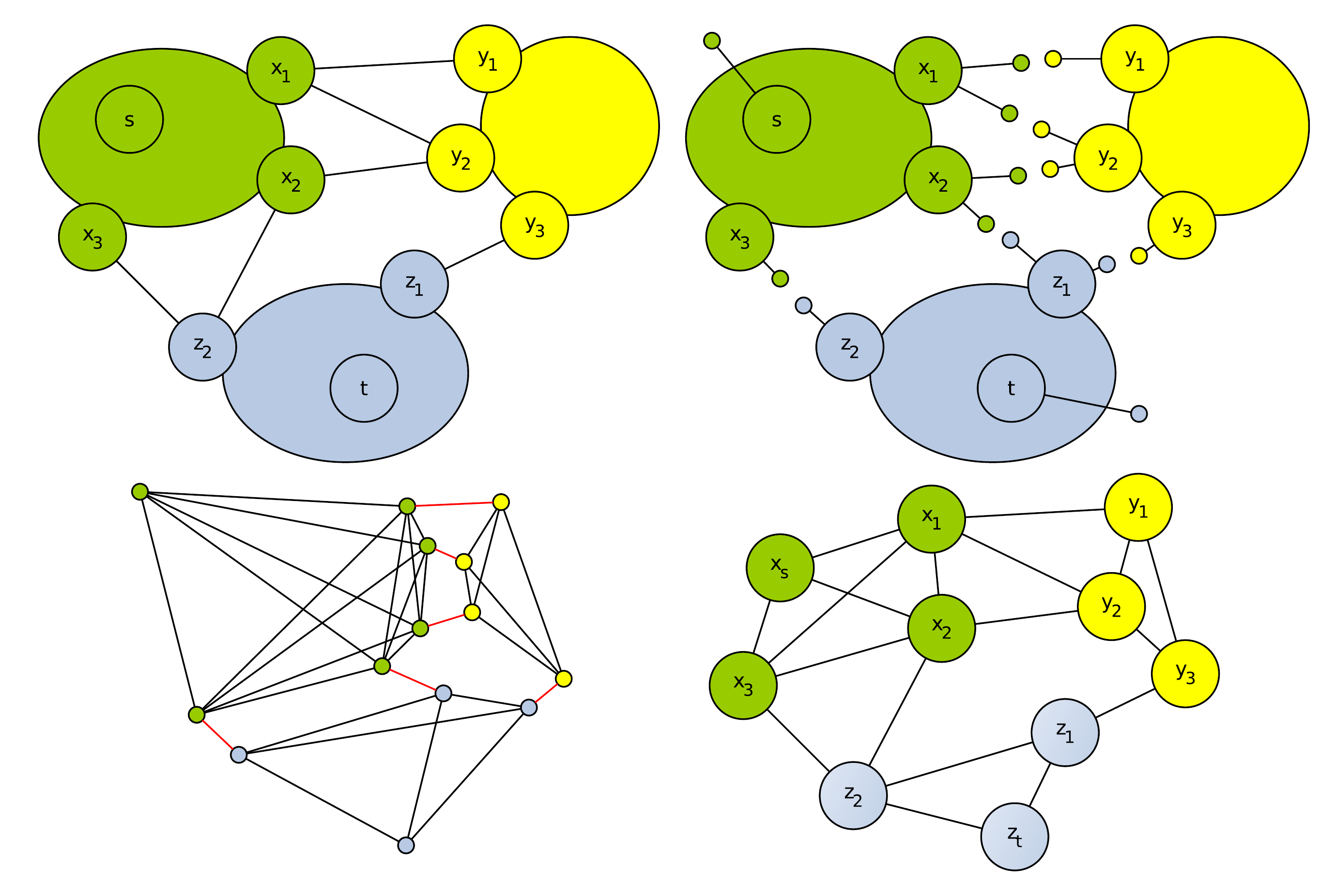}}
  \caption{An example illustrating the basic idea of the LPDP algorithm. 
  In the upper left corner we have a graph
  partitioned into three blocks indicated by the colors green, yellow and blue. The starting node is in the
  green block and the target in the blue block. The border nodes are named $x_i$, $y_i$ and $z_i$.
  In the upper right corner we see the three graphs created by removing the edges connecting the blocks and
  adding the auxiliary nodes (indicated as small circles). Finally, in the bottom we have the graphs used in
  the second stage of the algorithm -- combining paths. On the left side we see the simple version 
  and right side contains the improved version with fever auxiliary nodes and vertices.}
  \label{fig:example}
\end{figure}

We now explain our preprocessing routine. First of all, we partition a graph into a predefined number of blocks and modify the input instance.
Partitioning can be done using a graph partitioning algorithm, \eg KaHIP.
We then \emph{replace} every cut edge $e=\{x,y\}$, \ie an edge running between two blocks, by introducing two 
new vertices $v_e$,$v'_e$ and the edges $\{x,v_e\}$ and~$\{v'_e,y\}$.
One of these edges retains the weight of the original edge $e$, the other edge weight is set to zero. 
Additionally, we insert two new vertices. 
One is connected to the start vertex and the other one to the target vertex. In both cases, we use an edge having weight zero. 
All newly generated vertices are called \emph{auxiliary-vertices} throughout the rest of the section. 
Next, we compute subgraphs $G_{A} := (V_{A}, E_{A})$ where $V_{A}$ is the set vertices of block $A$ as well as all auxiliary-vertices connected to them, and $E_{A}$ are all edges that run between those vertices.
See Figure~\ref{fig:example} for an example.

Now observe the following property: for a longest simple path in $G$ the auxiliary-vertices function as entry and 
exit points for their block of the partition. That means a longest simple path from $s$ to $t$ can only 
enter and exit a block through the inserted auxiliary-vertices. For a block that does not contain $s$ or $t$,
every time the path enters a block, it also has to leave it again, connecting auxiliary-vertices in pairs of two. 
The sets of auxiliary-vertex-pairs $\mathcal{P}_{A}$ that we are interested in for any subset $s$ of the auxiliary-vertex-pairs for any block are equivalent to the matchings that 
exist for a clique-graph consisting of all auxiliary-vertices of that block. 
In other words, each endpoint can only appear once in a subset of pairs.
The pairs have to be connected by \emph{non-intersecting} simple paths. 
Our preprocessing algorithm computes the longest of these connections for each block and set of auxiliary-vertex-pairs with a modified 
version of exhaustive DFS executed on the respective subgraph, whose descriptions follows.

Our modified exhaustive DFS is executed multiple times each time using a different auxiliary-vertex as a root. The search algorithm divides 
its current search path into different path segments. It traverses the vertices of the graph as usual with the exception of the auxiliary-vertices. The first segment starts from the root auxiliary-vertex. 
The segment is completed once a different auxiliary-vertex is reached. 
When this happens, the algorithm starts a new segment by jumping to an other auxiliary-vertex
 and continues traversing the graph as before. 
This way each segment starts and ends in a auxiliary-vertex. The start- and endpoints of all segments 
resemble the auxiliary-vertex pairs $\mathcal{P}_{A}$. The best current result for each possible element in $\mathcal{P}_{A}$ is stored and updated if necessary every time a path~segment~is~completed. 
\csch{clarify whats actually contained in $\mathcal{P}_{A}$}

To avoid unnecessary traversal (due to symmetry) of the graph, a path segment is only allowed to end in a vertex 
having an id higher than its start vertex. Additionally a path segment can only start from an auxiliary-vertex, if this 
vertex is higher than all other starting vertices in the current search path.
\csch{TODO: unify vertices and vertices}

\subsubsection*{Combining Paths}
\label{sec:approach:step2}
After having performed preprocessing, we can now find the longest (simple) path of $G$ in an auxiliary graph that only 
contains auxiliary-vertices and edges between them. In this graph two auxiliary-vertices are connected by an edge 
if they belong to the same graph~$G_{A}$. Note that every block of the original partition is now 
represented by a clique of its auxiliary-vertices in our auxiliary graph. 
Additionally, the edges $\{v_e,v'_e\}$ get introduced for all edges $e$ that were replaced in the previous graph, where $v_e$ and $v'_e$ are the auxiliary-vertices that replaced $e$. These edges represent the connections between the blocks. 


In order to solve the longest path problem, we use another modified version of exhaustive DFS.
It starts from the vertex representing the start vertex. The algorithm creates a set of auxiliary-vertex-pairs $\mathcal{B}_{A} \in \mathcal{P}_{A}$ for every block $A$ from its search path. An edge $\{v,w\}$ is part of a block $A$ if both $v$ and $w$ are auxiliary-vertices of $A$. While this edge is part of the search path the pair $\{v,w\}$  is an element of $\mathcal{B}_{A}$. The pair $\{v,w\} \in \mathcal{B}_{A}$ represents a connection of the 
corresponding auxiliary-vertices of $v$ and $w$ in $G_{A}$ through a simple path. 
The simple paths of all these pairs in $\mathcal{B}_{A}$ cannot intersect with each other. 
The best possibility to do this and the combined length of these paths has already been calculated in the preprocessing phase.
We do the following in order to only receive valid auxiliary-vertex pairs $\mathcal{B}_{A}$ when trying to append new edges to the current search path:
first of all, for every block $A$ a solution for $\mathcal{B}_{A}$ has to exist. This can be looked up, since the best possible solutions have been calculated in the preprocessing step. Second, we have to search the graph in an alternating pattern between edges that are part of a block and edges that connect two blocks.
Otherwise $\{a,b\},\{b,c\} \in \mathcal{B}_{A}$ would be possible, which means that 
the two paths in $G_{A}$ would intersect.
Now every time the vertex representing of the original target-vertex has been found, the paths 
in $G_{A}$ for every $\mathcal{B}_{A}$ are looked up and their weight summed up. 
At the end the different $\mathcal{B}_{A}$ with the highest combined weight are returned. 
This weight is the weight/length of the longest simple path in $G$. The actual longest simple 
path can be constructed by looking up all the pre-calculated paths in $G_{A}$ for the given 
connections of its auxiliary-vertices $\mathcal{B}_{A}$. 

\subsubsection*{Improvement through Hierarchical Partitioning}
\label{sec:approach:improvement}
While preprocessing and computations can be done fast and even in parallel, 
the auxiliary graph has to be searched with a variant of the naive brute-force approach. 
We can avoid this by applying the same principles that we used to 
accelerate exhaustive~DFS~recursively. 

The \emph{main idea} here is to combine groups of blocks and the paths that have been calculated for each of the blocks into a single new coarser block. 
After assigning each vertex to its new/updated block we split the edges that connect different blocks as before. Note that the cliques representing the previous blocks stay intact. The auxiliary-vertices representing the start or target vertex is also connected to a new auxiliary-vertex as before.
The computation or combination of the paths is again done using the modified exhaustive DFS between the 
new block's auxiliary-vertices. In this way, we are actually doing a recursive call of our algorithm. 
In a sense, the partition of the original graph $G$ has to be hierarchical. 
Blocks are subsequently combined step by step into larger ones, until only one is left and a longest path of $G$ is calculated.

\subsubsection*{Improvement through Reducing the Number of Auxiliary Vertices}
Above we defined that every cut edge $e=\{x,y\}$ is replaced with two auxiliary-vertices and edges $\{x,v_e\}$ $\{v_e',y\}$. Let $\mathcal{A}_x := \{b_1,b_2,...\}$ be the set of all auxiliary-vertices that are connected to a vertex $x$. 
If we ignore the possibility of two elements of $\mathcal{A}_x$ connecting to each other ($b_i - x - b_j$), we see that a path segment of the current block connecting $b \in \mathcal{A}_x$ is also a valid path if we replace it with any other $b' \in \mathcal{A}_x$. 
This leads to an improved algorithm that only has a single auxiliary-vertex $b_x$ for every vertex. 
Hence, in contrast to the previous formulation, the weight of a cut edge $\{x,y\}$ does not get divided among the replacing edges $\{x,v_e\}, \{v_e',y\}$ as before. Instead the edge $\{v_e,v_e'\}$ that gets introduced in the auxiliary graph while combining the paths retains the weight of $\{x,y\}$. The weights of $\{x,b_x\}, \{b_y,y\}$ are set to zero. 
Note that in this variant of the algorithm, $b_x$ is the same for every other edge $\{x,z\}$ that is replaced, meaning $b_x$ is the only auxiliary-vertex~of~$x$.

In turn, the auxiliary graph cannot be traversed as before, since all edges between different blocks have weights associated with them. To obtain the correct result we sum up the weight of all weighted edges in the search path and add it to the length of the currently induced path. Additionally a single vertex can be connected to multiple different blocks. To still search through all possible valid paths we have to allow multiple consecutive edges in the search path that connects different blocks, compared to the purely alternating pattern from before. 
Entering and leaving a vertex $x$ through two such edges corresponds to a ($b_i - x - b_j$) connection mentioned above, which represents the usage of a single vertex $x$ in the block. 
It follows that the search path not only induces the set of auxiliary-vertex-pairs $\mathcal{B}_{A}$ for a block $A$, but also the set of ``excluded'' vertices $\mathcal{C}_{A}$. 
Hence, we have to find the best possibility to connect the pairs in $\mathcal{B}_{A}$ without using any vertices in $\mathcal{C}_{A}$. 
During preprocessing, we calculate all possibilities for each $\mathcal{B}_{A}$ and choose the best one. 
By looking at the unused vertices of a current solution, we get the best possibility for each $\mathcal{B}_{A}$ and $\mathcal{C}_{A}$. 
Note that any vertex in $\mathcal{C}_{A}$ would simply correspond to an additional pair $\{b_i,b_j\} \in \mathcal{B}_{A}$ in the previous formulation of~the~algorithm.
For an example of the improved auxiliary graph see Figure~\ref{fig:example}.

\section{Experimental Evaluation}
\label{s:experiments}
\paragraph*{Methodology} We have implemented the algorithm described above using C++ and compiled it using gcc 4.9.4 with full optimizations turned on (-O3 flag). Our implementation is freely available in the Karlsruhe Longest Paths package (KaLP) under the GNU~GPL~v2.0 license~\cite{kaLPHomePage}. 
We use multiple implementations provided by Stein \etal \cite{stern} for comparison: $\textbf{Exhaustive DFS}$ is the naive brute-force approach as well as the $\textbf{A*}$ algorithm and the $\textbf{DFBnB}$ solver.
We run each algorithm and input pair with a time limit of one hour.
Experiments were run on a machine that is equipped with two Intel® Xeon® Processors X5355 (2.66 GHz with 4 cores) and 24 GB RAM. 

We present multiple kinds of data:
first and foremost, we use \emph{cactus plots} in which the number of problems is plotted against the running time. The plot shows the runtimes achieved by the algorithm on each problem. The running times are sorted in ascending order for each algorithm. The point ($x$, $t$) on a curve means that the $x$th fastest solved problem was solved in $t$ seconds. Problems that were not solved within the time limit are not shown. In addition we utilize tables reporting the number of solved problems as well as \emph{scatter plots} to compare running times of two different solvers $\mathcal{A},~\mathcal{B}$ by plotting points $(t_\mathcal{A}, t_\mathcal{B})$ for each instance.

\paragraph*{Benchmark Problems}
We mainly use instances similar to the ones that have been used in previous work by Stein~\etal\cite{stern}, \ie based on mazes in grids as well as the road network of New York. Additionally we use subgraphs of a word association graph~\cite{webgraphWS}. The graph describes the results of an experiment of free word association performed by more than 6000 participants. Vertices correspond to words and arcs represent a cue-target pair.

\begin{wrapfigure}{r}{5cm}
\centering
\vspace*{-.5cm}
\begin{tikzpicture}

\draw[step=5mm,gray,very thin] (0,0) grid (5,5);
\fill[black] (1.0,0.0) rectangle (1.5,0.5);
\fill[black] (1.5,0.0) rectangle (2.0,0.5);
\fill[black] (2.0,0.0) rectangle (2.5,0.5);
\fill[black] (3.5,0.0) rectangle (4.0,0.5);
\fill[black] (4.0,0.0) rectangle (4.5,0.5);

\fill[black] (0.5,0.5) rectangle (1.0,1.0);
\fill[black] (1.0,0.5) rectangle (1.5,1.0);
\fill[black] (1.5,0.5) rectangle (2.0,1.0);
\fill[black] (2.5,0.5) rectangle (3.0,1.0);
\fill[black] (3.5,0.5) rectangle (4.0,1.0);
\fill[black] (4.0,0.5) rectangle (4.5,1.0);

\fill[black] (0.0,1.0) rectangle (0.5,1.5);
\fill[black] (1.0,1.0) rectangle (1.5,1.5);
\fill[black] (2.0,1.0) rectangle (2.5,1.5);
\fill[black] (4.0,1.0) rectangle (4.5,1.5);

\fill[black] (1.0,1.5) rectangle (1.5,2.0);
\fill[black] (1.5,1.5) rectangle (2.0,2.0);
\fill[black] (2.5,1.5) rectangle (3.0,2.0);
\fill[black] (4.0,1.5) rectangle (4.5,2.0);

\fill[black] (0.5,2.0) rectangle (1.0,2.5);
\fill[black] (2.0,2.0) rectangle (2.5,2.5);
\fill[black] (3.0,2.0) rectangle (3.5,2.5);
\fill[black] (3.5,2.0) rectangle (4.0,2.5);
\fill[black] (4.0,2.0) rectangle (4.5,2.5);

\fill[black] (0.0,2.5) rectangle (0.5,3.0);
\fill[black] (0.5,2.5) rectangle (1.0,3.0);
\fill[black] (2.5,2.5) rectangle (3.0,3.0);

\fill[black] (2.5,3.0) rectangle (3.0,3.5);

\fill[black] (2.0,3.5) rectangle (2.5,4.0);
\fill[black] (4.0,3.5) rectangle (4.5,4.0);
\fill[black] (4.5,3.5) rectangle (5.0,4.0);

\fill[black] (0.5,4.0) rectangle (1.0,4.5);
\fill[black] (1.5,4.0) rectangle (2.0,4.5);
\fill[black] (3.0,4.0) rectangle (3.5,4.5);
\fill[black] (3.5,4.0) rectangle (4.0,4.5);
\fill[black] (4.0,4.0) rectangle (4.5,4.5);

\fill[black] (0.5,4.5) rectangle (1.0,5.0);
\fill[black] (2.5,4.5) rectangle (3.0,5.0);
\fill[black] (4.0,4.5) rectangle (4.5,5.0);

\node[draw=none] at (0.25,4.75) {s};
\node[draw=none] at (4.75,0.25) {t};

\draw[line width=0.5mm, red] (0.25,4.60)
-- (0.25,3.25)
-- (1.25,3.25)
-- (1.25,2.25)
-- (1.75,2.25)
-- (1.75,2.75)
-- (2.25,2.75)
-- (2.25,3.25)
-- (1.75,3.25)
-- (1.75,3.75)
-- (1.25,3.75)
-- (1.25,4.75)
-- (2.25,4.75)
-- (2.25,4.25)
-- (2.75,4.25)
-- (2.75,3.75)
-- (3.75,3.75)
-- (3.75,3.25)
-- (3.25,3.25)
-- (3.25,2.75)
-- (4.25,2.75)
-- (4.25,3.25)
-- (4.75,3.25)
-- (4.75,0.40);

\end{tikzpicture}
\caption{An example maze with obstacles and the longest $s$-$t$ path.}
\label{fig:mazeexample}
\vspace*{-.25cm}
\end{wrapfigure}
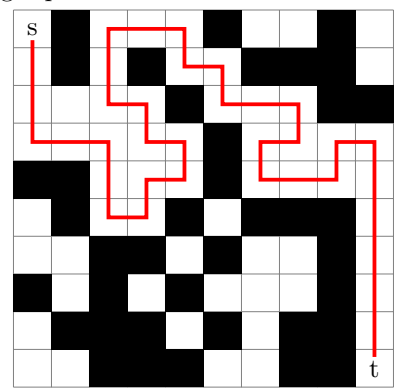
The first set of instances is generated by using mazes in $N\times N$ grids of square fields with a given start and target field. 
One can move to adjacent fields horizontally or vertically but only if the field is not an obstacle.
The goal is to find the longest simple path from the start field to the target field. 
We represent the grids as graphs: for every free field we insert a vertex and we add an edge with weight 1 between any two vertices, whose fields are horizontally or vertically adjacent to each other.
We generate the grids as Stein~\etal\cite{stern}: the top left and bottom right field are the start and target fields. Random fields of the grid are consecutively made into obstacles until a certain percentage $x\in\{30\%, 40\%\}$ of all fields is filled. Afterwards a path between the start and target is searched for to make sure that a solution of the longest~path~problem~exists.
An example maze is shown in Figure~\ref{fig:mazeexample}.
The sizes of the used mazes range from 10x10 up to 120x120.

The second and third set of instances are subgraphs of the road network of New York~\cite{dimacschallenroad} as well as subgraphs of the word association graph~\cite{webgraphWS}, respectively. A subgraph is extracted as follows: we start a breadth-first search from a random vertex of the network and stop it when a certain number of vertices is reached. The vertices touched by the breadth-first search induce the instance. 
One of the touched vertices is randomly chosen as the target-vertex.

\paragraph*{Experimental Results}
We now compare A*, DFBnB, and exhDFS presented by Stein~\etal\cite{stern} to
our algorithm LPDP using two configurations. Our configurations differ in the amount of time that we spent on partitioning the input instance. We use either the eco/default configuration of KaFFPa (\emph{LPDPe}), which is a good trade off between solution quality and running time, or the strong configuration of KaFFPaE which aims at better partitions while investing more time for partitioning (\emph{LPDPs}). In the latter case, we set the amount of block imbalance to 10\%.
Note that LPDPs spends much more time in the graph partitioning phase of the algorithm than LDPDe. 
All results reporting running time in this paper include the time spent~for~partitioning.
\begin{figure}
  \centerline{\includegraphics[width=0.8\linewidth]{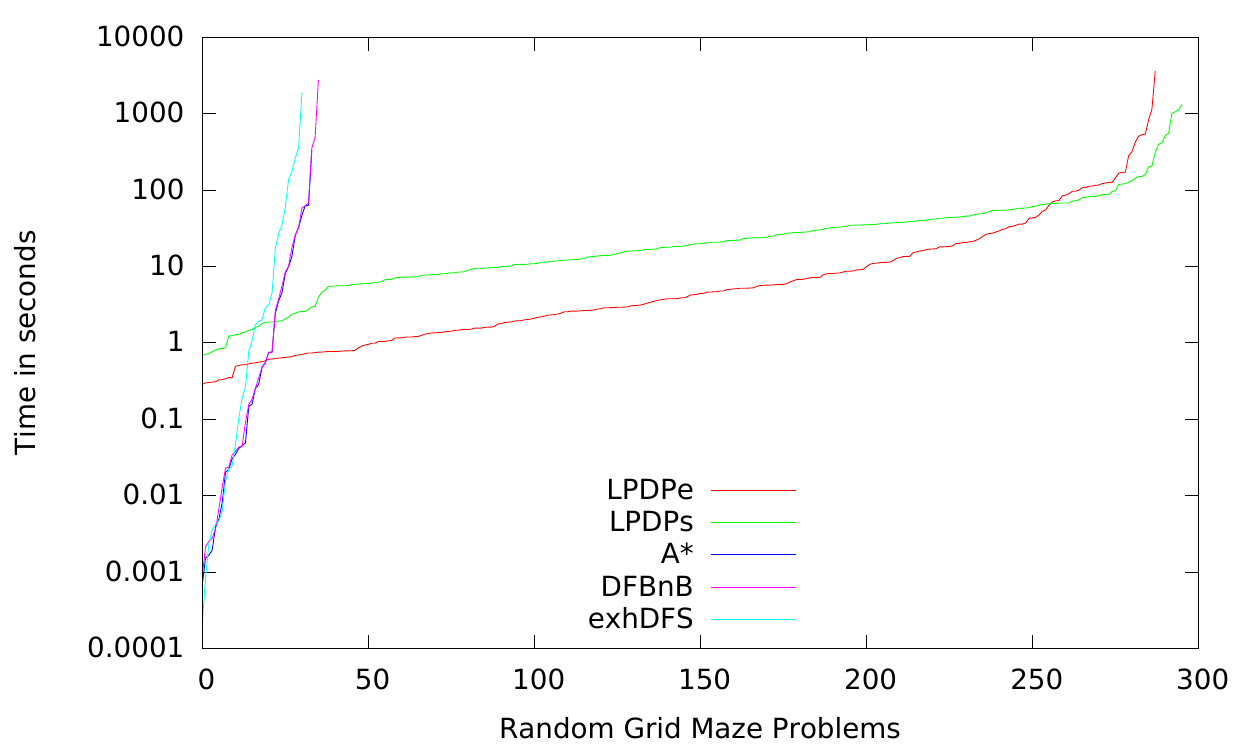}}
  \centerline{\includegraphics[width=0.8\linewidth]{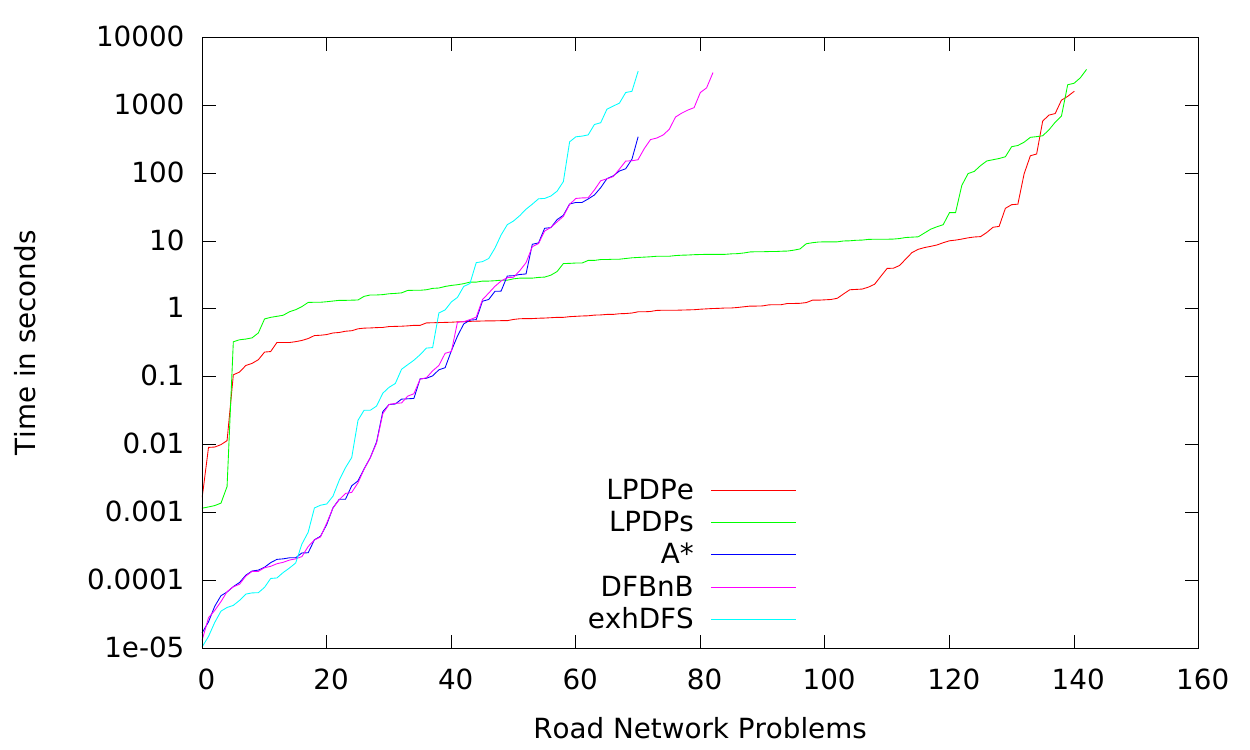}}
  \centerline{\includegraphics[width=0.8\linewidth]{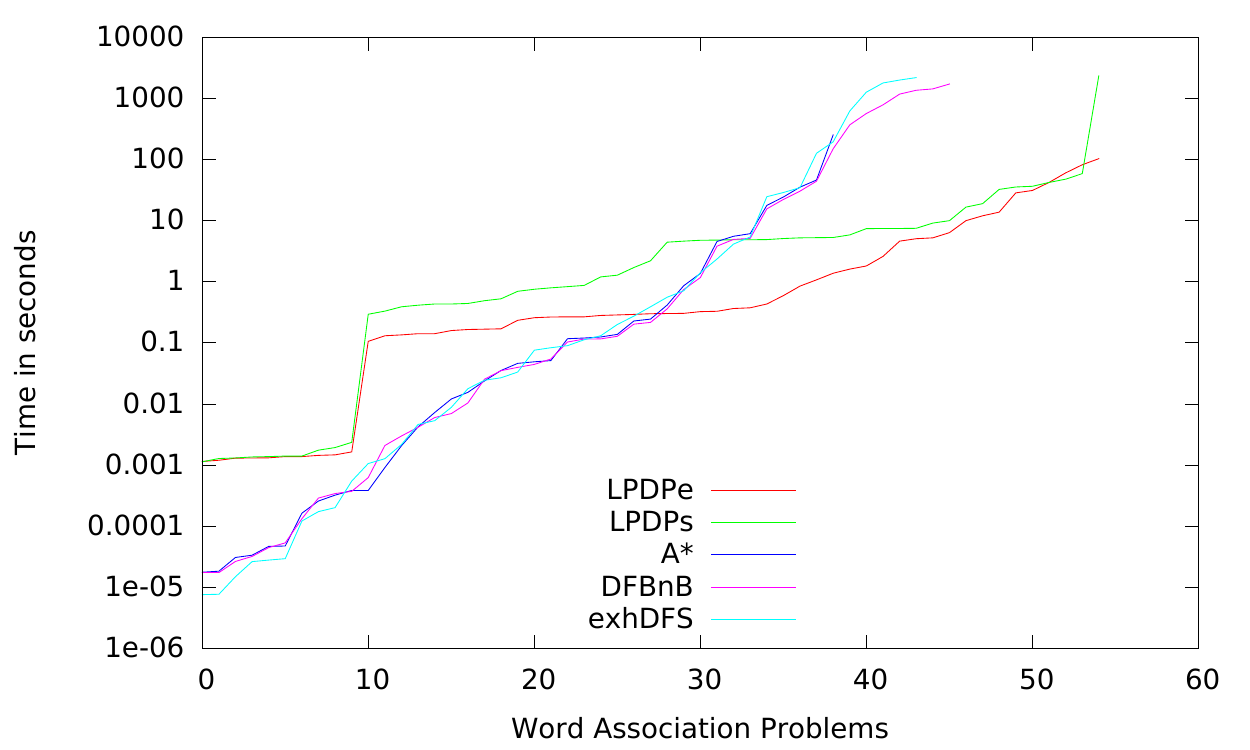}}
  \caption{Cactus plots for the three kinds of benchmark problems comparing the previous algorithms to
  LPDP which three different partitioning configurations. The running times include time spent on
  partitioning for the LPDP variants.}
  \label{fig:cactuses}
\end{figure}

\begin{figure}
  \centerline{\includegraphics[width=0.9\linewidth]{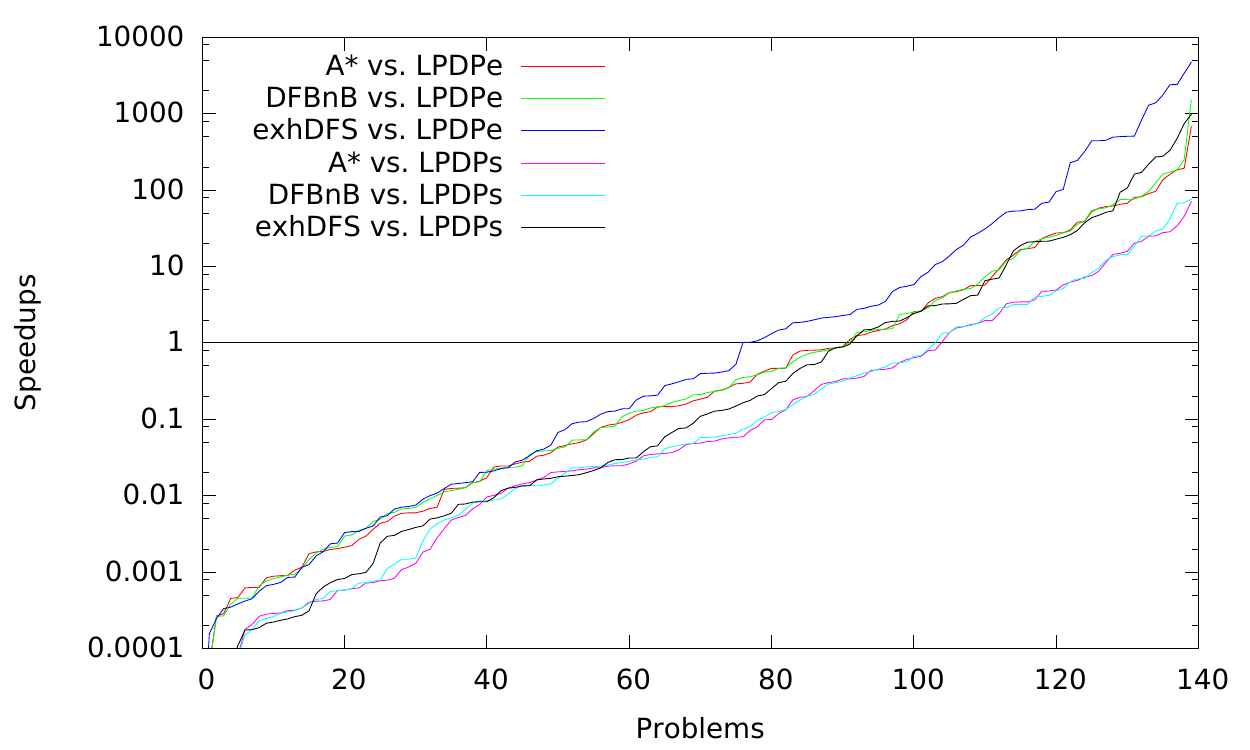}}
  \caption{Speedup of LPDPe and LPDPs in relation to previous LP algorithms on 
  problems that were solved within the time limit by each of the five tested algorithms.}
  \label{fig:speedup}
\end{figure}

\begin{figure}
  \centerline{\includegraphics[width=0.9\linewidth]{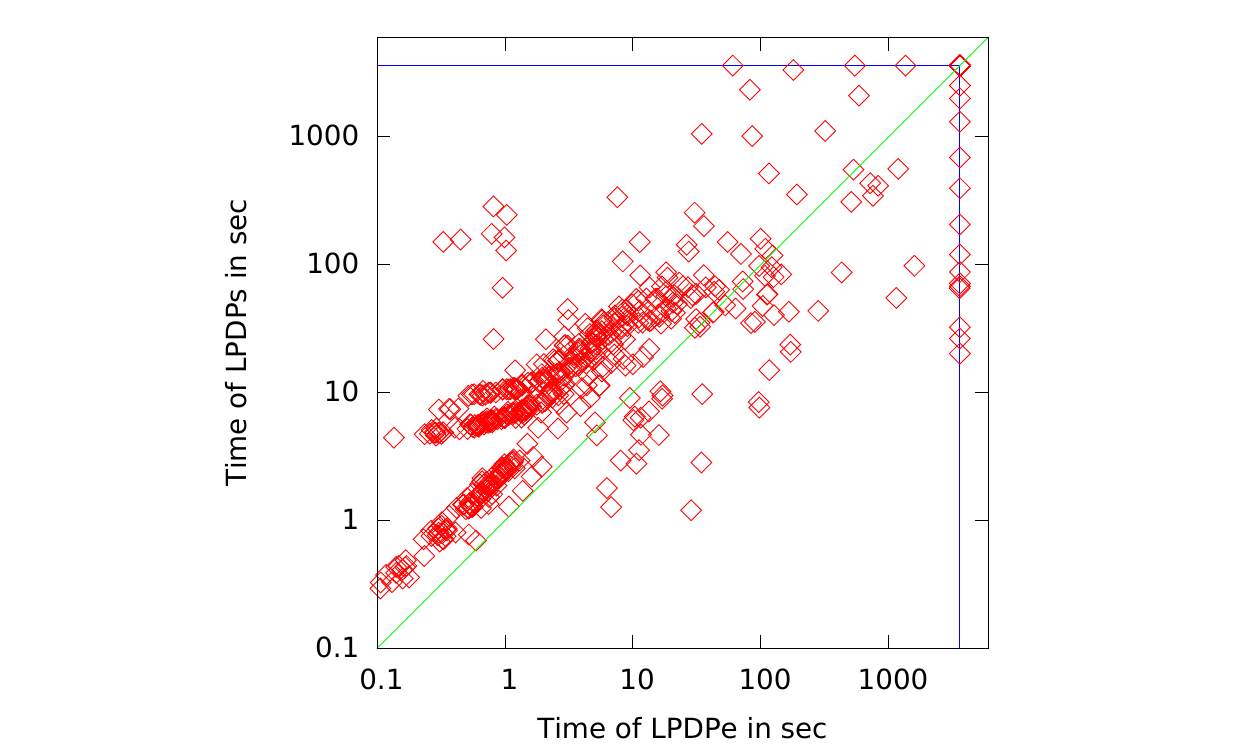}}
  \caption{A scatter plot comparing running times of LPDPe  and LPDPs  
  on the entire benchmark set.
  Points above (below) the green line represent instances where LPDPe (LPDPs) 
  was faster.
  Points on the right (top) blue line represent instances that were not solved within the time limit of
  one hour by LPDPe (LPDPs).}
  \label{fig:scatter}
\end{figure}

\begin{table}[t]
\begin{center}
\begin{tabular}{l||rrrr} 
  \multirow{2}{*}{Solver} & \multicolumn{4}{c}{Number of Solved Instances}\\ 
  & Grid Mazes & Road Network & Word Assoc.& Total\\ 
 \hline
 A*              & 34 & 71 & 39 &  144\\ 
 DFBnB           & 36 & 83 & 46 &  165\\ 
 Exhaustive DFS & 31 & 71 & 44 &  146\\ 
 \hline
 LPDPe                  & 288 & 141 & 55 & 484 \\ 
 LPDPs & 296 & 143 & 55 & 494 \\ 
\end{tabular}
\end{center}
\caption{The number of instances solved within the time limit of one hour 
by the tested solver configurations for each collection of benchmark problems and in total.}
\label{tab:solved}
\end{table}

Figures~\ref{fig:cactuses}--\ref{fig:scatter} and Table~\ref{tab:solved} summarize the results of our experiments.
It is apparent from the cactus plots in Figure~\ref{fig:cactuses} that both configurations of LPDP significantly outperform the previous algorithms for
each kind of tested benchmark except for very easy problems. These problems are typically solved under a few seconds by any of the
algorithms. In these cases, most of the time of our algorithm is spent in the partitioning phase. Moreover, our LPDP algorithms can solve significantly more problems, which can be seen in the cactus
plots as well as in Table~\ref{tab:solved}.

There are 140 problem instances that were solved by all five solvers within the time limit. 
In Figure~\ref{fig:speedup} we provide the speedup of LPDPe and LPDPs against the three 
original LP algorithms.
For most of these instances the speedup is actually below 1, but from our data we know, that this
happens only for easy problems (solvable within a couple of seconds by each solver). The slowdown on these easy instances is due to
the overhead caused by partitioning. Nevertheless, the average speedups are still above 1:
17.45, 24.44 and 166.01 for LPDPe vs. A*, DFBnB, and exhDFS respectively and 3.24, 3.83 and 31.22
for LPDPs vs. A*, DFBnB, and exhDFS respectively.

The differences in running time are highest for the grid maze instances and lowest for word association graph problems.
We believe this is due to the structure of these graphs, in particular, how well they can be partitioned
to loosely connected subgraphs. Our algorithm excels on problems that can be successfully partitioned but
is competitive on all~kinds~of~graphs.

As of evaluating our algorithm with different partitioning configurations, we see that spending extra time
on partitioning to get better solutions pays off. In particular, LPDPs is able to solve more instances. Especially if the instance appears to be hard it is worth while to invest more time in partitioning. Additionally, this depends
on how well the graphs can be partitioned (highest for grid mazes, smallest for word association). 

Looking at the scatter plot in Figure~\ref{fig:scatter}, we can see that LPDPe is faster for most of the
instances but has significantly more unsolved instances. Nevertheless, there are 
three instances that are solved by LDPDe and not by LPDPs. Each of these three instances come from
a different benchmark set. This shows that spending more effort on the partitioning does not necessarily
increase the number of solved instances.

\section{Conclusion}
\label{s:conclusion}
We presented an optimal algorithm for the longest path problem in undirected graphs which is based on dynamic programming.
Experiments show that our new algorithm is faster for nontrivial problems than the previous optimal algorithms and can solve significantly more benchmark instances if a time limit per instance is given. 
Important future work includes parallelization of our algorithm to improve the solver speed even further. 




\bibliography{citations,phdthesiscs}


\end{document}